\documentclass[
 aip,
 jcp,
 amsmath,amssymb,
 reprint,%
]{revtex4-1}
\usepackage{graphicx}
\usepackage{dcolumn}
\usepackage{bm}
\usepackage[utf8]{inputenc}
\usepackage[T1]{fontenc}
\usepackage{mathptmx}
\usepackage{float}

\begin{document}

\title{High harmonic generation spectroscopy via orbital angular momentum}
\author{Jan Tro\ss}
\affiliation{James R. Macdonald Laboratory, Department of Physics, Kansas State University, Manhattan, KS, USA}
 \author{Carlos Trallero}%
 \email{carlos.trallero@uconn.edu}
\affiliation{Department of Physics, University of Connecticut, Storrs, CT, USA}
\date{\today}
\begin{abstract}
We present an experimental technique using orbital angular momentum (OAM) in a fundamental laser field to drive High Harmonic Generation (HHG). The mixing of beams with different OAM allows to generate two laser foci tightly spaced to study the phase and amplitude of HHG produced in diatomic nitrogen. Nitrogen is used as a well studied system to show the quality of OAM based HHG interferometry.
\end{abstract}
\maketitle

\section{Introduction \label{sec:intro}}

Higher-order harmonic generation (HHG) \cite{corkum1993PRL,krause1992PRL,lhullier1993PRL,mcpherson1987josab,ferray1988jopb} has been shown to be a powerful spectroscopic technique\cite{baker2006science,itatani2004nature,Lein2002PRA,velotta2001PRL,kanai2005nature,zhou2008PRL,mairesse2005PRA,smirnova2009nature} and has been used for twenty years to push atomic and molecular physics to the attosecond regime, allowing molecular structures \cite{baker2006science} and charge migrations in the attosecond regime \cite{kraus2015science}.\newline
Harnessing the coherence of light and specifically of the emitted higher harmonics, this letter shows another approach for measuring complex signals $S(\omega,t)=|S(\omega,t)|e^{i\phi(\omega,t)}$ in the extreme ultraviolet (XUV) from HHG, which then can be used to study structural details of diatomic nitrogen. More specifically, XUV light emitted through HHG holds complex-valued information about the photorecombination dipole matrix element $d(\omega,\theta)=|d(\omega,\theta)|e^{i\phi(\omega,\theta)}$, where the recombination process is understood as the time reversal of photoionization \cite{Shiner2011NPhys}. The time reversal relationship between photoionization and photorecombination is better understood through the principle of detailed balanced \cite{LL3}, a principle that has been already applied to the description of the generation of harmonics \cite{Frolov09PRL, frolov2016PRA}. The relevance of the photoionization dipole lies, of course, in the fact that it has been widely used by chemists and physicists as a tool to investigate atomic and molecular structures. However, in the weak field regime of photoionization, phase information about the transitions cannot be easily retrieved. \\
In contrast, HHG offers direct access to the amplitude and phase of the photoionization dipole. Calculations of the photorecombination dipole aided by experiments, have been presented in nitrogen \cite{jin2012PRA} and show energy and angle dependent variations. A shape resonance at 30~eV shows dramatic enhancement of the cross section and a variation of the phase as a function of energy. Measurements of the angle and energy dependent cross section have been measured through other techniques such as the reconstruction of attosecond beating by interference of two-photon transitions (RABBITT)\cite{paul2001science, Muller2002APA}.  In N$_2$, phase information has been investigated \cite{Ferre2015NatComm} in RABBITT and through interferometric studies \cite{lock2009chemphys,camper2015photonics} with separate beam paths, as well.\\
However, even with table-top techniques such as RABBITT, extracting these features is experimentally challenging and requires a stable interferometric setup. This is were our proposed experimental techniques comes in play. Using a programmable spatial light modulator (SLM), we can build a stable, reliable common-path interferometer that is insensitive to instabilities in experimental setups. \newline
We use the well documented features of diatomic nitrogen to benchmark our technique of using beams with orbital angular momentum, while studying the angle dependence of individual harmonics with molecular alignment. In nitrogen, similar measurements have already been performed and showed reliable features with different details:
In a 2015 publication Camper et al \cite{camper2015photonics} used a binary phase mask to study the change in phase of harmonics, as the molecules rotate in time and shows for harmonics 9 to 17 an oscillatory behavior of the phase, while a measurement by Lock et al \cite{lock2009chemphys} showed no change in phase for harmonic 19. The key difference in the experiments is the way two intense laser fields are produced. We will show results that also extend the measurements to the molecular frame. Further, publications \cite{mcfarland2008science,rupenyan2012PRL} have also shown the impact of multiple orbitals or the dominance of single orbitals.
When using molecules as targets, the electronic configuration can be described with molecular orbitals \cite{fukui1952JCP,slater1951physrev}, where the electron in the highest occupied molecular orbital (HOMO, $3\sigma_g$) has the smallest binding potential with the molecule and the next lower lying orbital is defined as the HOMO-1 ($1\pi_u$). In nitrogen, these two orbitals have ionization potentials of 15.6~eV for the HOMO and 16.9~eV for the HOMO-1. These binding potentials are very close to each other and ionization rates can become similar which needs to be taken into account when using HHG as a spectroscopic tool \cite{Tros2017PRA}.

\section{Experimental setup\label{sec:expt}}

\begin{figure}
 \centering
        \includegraphics[width=\columnwidth]{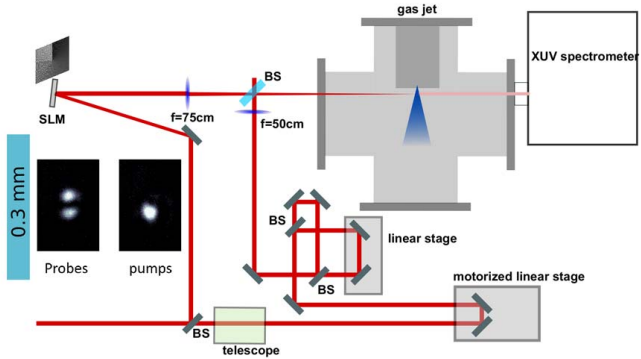}
    \caption[Experimental setup for nitrogen measurements]{Experimental setup with an unbalanced Michelson interferometer to separate the input into one probe and two pump arms. The probe and pump beams are focused into a pulsed gas jet, in which high harmonics are generated. A spectrometer is used to measure the light's amplitude and phase. In the left inset, pump and probe foci are shown in respect to the nozzle diameter of 300$\mu{m}$, preparing the pulsed gas sample before it interacts with the laser focus after 3~mm of travel and expansion.}
    \label{fig:n2_setup}
\end{figure}
We use a pump-probe scheme, where the pump pulse is inducing molecular alignment and the probe is driving high harmonic generation, which is used as our spectroscopic tool. The relative delay between pump and probe pulses is controlled by changing the optical beam path of the pump pulses by moving a mirror on a motorized linear stage. The experimental setup is shown in Figure \ref{fig:n2_setup}. In this setup, the probe arm is reflected off a beam splitter and is shaped by a spatial light modulator (SLM) to form a two source focus spot. The probe is focused by a f=75~cm lens into the gas jet and harmonics are generated by both focus spots. As the harmonics travel into the far field, harmonics from the two sources interfere and form a fringe pattern on the detector, which allows us to detect the relative phase between the harmonic sources and their intensity. \newline
The pump arm is transmitted through the first beam splitter and contains initially 80\% of the total pulse energy. After shrinking the beam by a factor of two with a telescope, the beam is split again by a set of two beam splitters in another interferometer inside the pump arm. By controlling the relative delay between the two pump pulses, we can tune the interaction of the pumps upon the molecular target and enhance the molecular alignment\cite{bisgaard2004PRA,cryan2009PRA} or potentially orient molecules in space \cite{kraus2014PRL,ren2014pra}. The pump beams are focused with a separate lens of f=50~cm into the center of the gas jet as seen in Figure \ref{fig:n2_setup}. The pump and probe beams are recombined with a beam splitter. The focus of the pump beam is placed at the center of the jet and is overlapped with one of the two probe beams. \newline
The spot size of the pumps are measured to be 68~$\mu{m}$ by 83~$\mu{m}$ and hold a pulse energy of 200~$\mu{J}$ with a pulse duration of 100~fs, measured by a cross correlation between the measured probe of 30~fs and the pump beams, which results in an experimentally determined pump intensity of 21~$TW/cm^2$ and a probe intensity of 110~$TW/cm^2$. The overlap between pump and probe is optimized on the live harmonic signal and we check that only one of the two sources is interacting with the pump beams (by using different sets of phase masks as highlighted in the later sections). Switching between masks is also used to find an optimum overlap between the gas jet and the two probe beams. The gas jet \cite{irimia2009revsci} has a nozzle of 300$\mu{m}$ after which the gas expands over 3~mm before able to interact with laser beams. The vertical offset is adjusted with a micrometer on the three dimensional manipulator, holding the jet. In the data acquisition, amplitude and phase of the harmonics are collected as a function of delay between pump and probe in step sizes of 40.04~fs. This step size is a sufficient sampling in time to capture smallest features in the revival structure from nitrogen, with a rotational period of 8.3~ps. In the later experimental data, we can see smaller features with periods of 240-400~fs, that can be resolved with the given step size. \\
As briefly mentioned before, a SLM is implemented to create a two-foci intensity distribution that in turn produces an interferometric pattern of the XUV pulses in the far field. Instead of using phase patterns that break the spatial symmetry\cite{camper2015photonics,zhou2008PRL} of the incoming beam, orbital angular momentum (OAM) is applied to generate two foci, imprinted by the whole surface of the SLM. The benefit of this approach is that the entire incoming beam will be affected by the phase distribution at the SLM which translates into two beams that are indistinguishable up to millimeters away from the interaction region. As a result, the XUV interferometer is completely self-referencing and insensitive to air currents and optics imperfections since both light paths have the exact same optical path.

\section{Details of Spatial Light Shaping}

For completeness we include details of the Fourier optics calculation that demonstrates the generation and control of two foci. Details of such methods can be found elsewhere \cite{voelz2011spie}. When monochromatic light with a spatial profile of $U(\eta,\nu)$ propagates from the source plane $(\eta,\nu)$ to the observation plane $(x,y))$ ,we can formulate the propagation with an impulse function $h$, that can be simplified to a simple propagator $h=\frac{e^{ikr}}{r}$. The field $U'(x,y)$ in the observation plane can be predicted using the first Rayleigh-Sommerfeld diffraction solution,
\begin{equation}
U'(x,y)=\frac{z}{i\lambda}\int\int U(\eta,\nu)\frac{exp(ikr)}{r^2}d\eta d\nu,
\label{eq:Rayleighsommerfeld}
\end{equation}
with $\lambda$ the wavelength, $k$ the wave number and $z$ the distance between the source and the observation plane.
In Equation \ref{eq:Rayleighsommerfeld}, $r$ is the distance between a position on the source plane and a position on the observation plane, $r=\sqrt{z^2+(x-\eta)^2+(y-\nu)^2}$. Different approximations for the square root term result in the Fresnel (and Fraunhofer) diffraction formulas, giving solutions to the diffraction through an aperture with the characteristic size $a$ \cite{voelz2011spie}. We choose the Fraunhofer approximation based on the Fresnel number $N=\frac{a^2}{L\lambda}$, with $L$ the distance to the observation plane (focus) from the aperture (SLM). Equation \ref{eq:Rayleighsommerfeld} can be used to calculate the focal spot of an arbitrary spatially-shaped pulse or, in our case, beams that have the purpose of generating two intense spots, to generate harmonics. \\
While the generation of harmonics driven by beams with OAM has been demonstrated \cite{Gariepy2014PRL}, exploring multi-foci experiments with OAM for HHG has not been demonstrated. In a previous publication we used beams with superposed OAM for micro-machining purposes \cite{Yu2016AppSurfSc} proving that intense multi-foci femtosecond beams with OAM can be generated.

\subsection{Orbital angular momentum}

\begin{figure}
\setlength{\lineskip}{0pt}
 \centering
  \includegraphics[width=90mm]{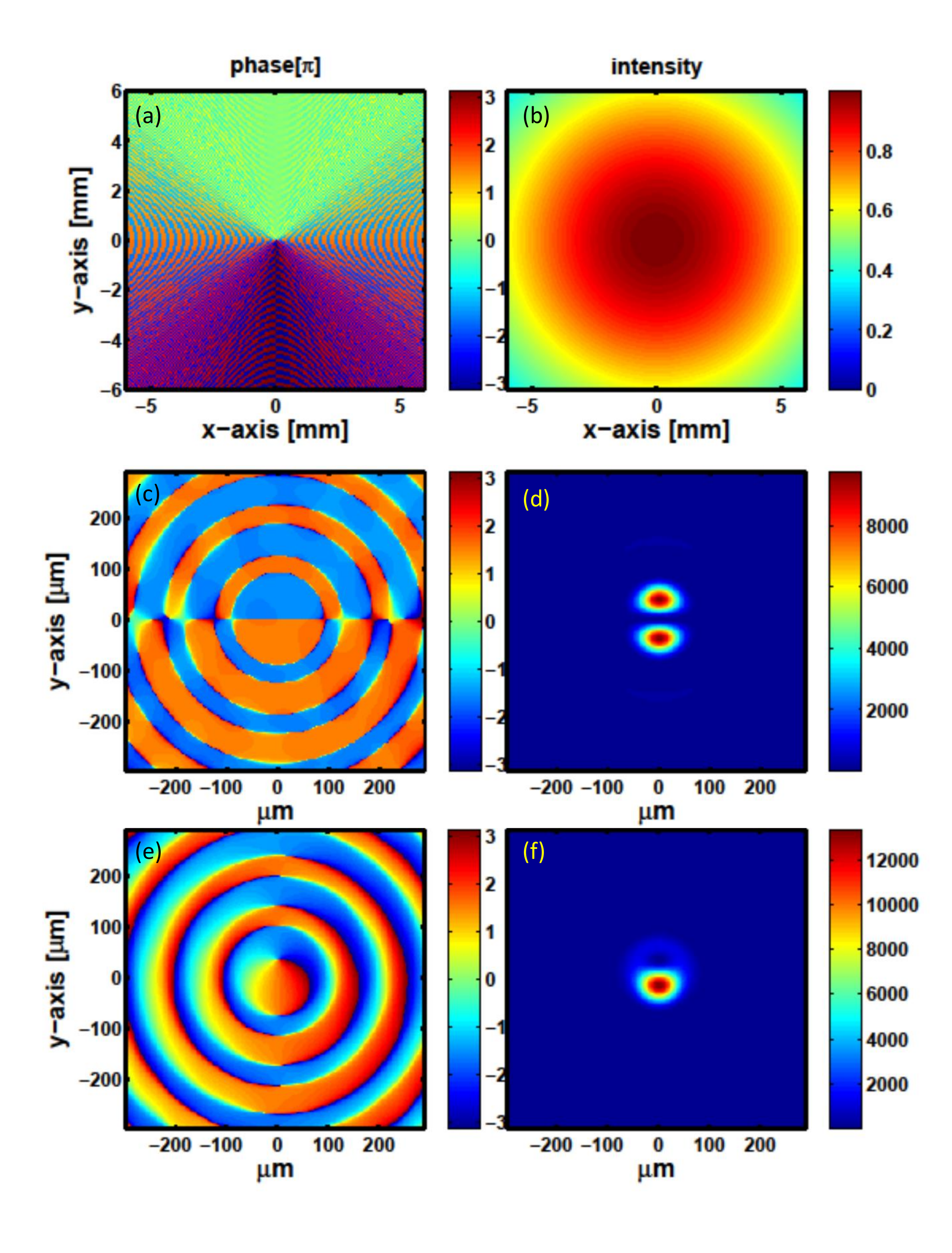}
     \caption[Spatial profile in intensity and phase for a phase profile with OAM]{Panel (a) and (b):Spatial phase and intensity distribution respectively in the source plane (SLM) after manipulation of the phase with OAM $l=1,-1$. (c) spatial phase and (d)  spatial intensity distribution at the focus of a 75~cm lens with a phase mask on source plane shown in (a) as given by Eq. \ref{eq:OAM_phase}. (e) phase and (f) intensity spatial distribution at the focus of a 75~cm lens after manipulation of the phase with OAM $l=1,0$.}
    \label{fig:input_LG}
\end{figure}

Using the aspect of OAM to shape the focus of our Gaussian beam into two separate foci, is the key to a stable interferometer. Throughout the paper we use the definition of Laguerre Gaussian (LG) beams as in \cite{allen1992PRA}. LG beams are an equivalent description of transverse electromagnetic modes, but in cylindrical rather than Cartesian coordinates and these modes carry so called orbital angular momentum. A strict definition of LG modes requires Laguerre polynomials and different normalization constants. However, we use a Gaussian mode that has OAM and shares in this regard similarities with LG beams.\\
We write a Gaussian beam as,
\begin{equation}
U(\rho,z)=E_0\frac{w_0}{w(z)^2}\exp\left[-\frac{\rho^2}{w(z)^2}-i(kz+k\frac{\rho^2}{2R(z)}-\Phi(z)\right],
\end{equation}
with $E_0$ the peak field strength, $w_0$ the initial beam waist, $w(z)$ the beam waist at position $z$, $\rho$ the radius, $R(z)$ the radius of curvature of the wavefront and $\Phi(z)$ the Gouy phase. To this Gaussian distribution, the SLM adds a phase term that carries OAM, $exp(il\theta)$
\begin{equation}
U^*(\rho,\theta,z)=U(\rho,z)exp(il\theta).
\label{eq:OAM_phase}
\end{equation}
Experimentally the SLM generates beams with $l=1$ and beams with $l=-1$,
resulting in a phase difference between the two different LG modes of $\Delta\Phi(\theta)=2\theta$. The phase mask and intensity profile used to simulate our beam propagation is shown in Fig. \ref{fig:input_LG} (a) and (b) respectively.
Generation of two superposed OAM beams with $l=\pm1$ can be thought of as two identical independent waves with different phases, $U_1=|U|e^{i\phi_1}$ and $U_2=|U|e^{i\phi_2}$ that when overlapped in space generate a space-dependent interference profile given by,
\begin{equation}
I=2U^2(1+\cos\Delta\phi),
\label{eq:inter}
\end{equation}
with $\Delta\phi=\phi_2-\phi_1$. The simulated intensity distribution at the focus of a $75~cm$ focal length lens has phase and intensity profiles shown in Fig. \ref{fig:input_LG} (c) and (d) respectively. Constructive interference occurs at integer multiples of $\Delta\phi=2l\pi$ or at the maximum two lobes in Fig.\ref{fig:input_LG} (d). At values of $\Delta\phi=(2l+1)\pi$ the two light beams interfere destructively which shows up as a minimum in the same figure. This beam profile is similar to what has been previously reported in \cite{bourouis1997JMO,passilly2005JOSA,smith2005JPB,meyrath2005OE} where a single Gaussian TEM$_{00}$ mode is manipulated by inserting a phase plate with a phase difference of $\pi$ between the upper and lower halves, resulting in two beams parallel to the optical axis or a beam similar to a quasi-Hermite Gaussian of order TEM$_{01}$, but only through the alternating sampling of OAM with opposing sign, can we achieve a stable mode. Inserting a phase plate of two zones does not guarantee a stable mode, as beam pointing instabilities would result in a changing intensity ratio and interference condition.

The SLM allows yet finer control over the spatial profile of the focused beams. For example, often one desires to have a single beam experiment to serve as reference. In such cases the one focus should be identical to a single focus of the multi-beam experiment. To achieve this, we generate a superposed beam with OAM l=0,1 (or -1). To understand how this approach works Eqs. (\ref{eq:OAM_phase}) and (\ref{eq:inter}) are used. At the focal plane, the phase difference between the beam with $l=0$ and the beam with $l=1$ is, 
\begin{equation}
\Delta\phi(\theta)=\phi_1-l\theta-\phi_2.
\label{eq:single_lg}
\end{equation}
where the beam with $l=0$ is again a pure Gaussian mode. For $l=1$  there is only one possible value for destructive interference and thus a single mode focus is created. This is clear in Fig. \ref{fig:input_LG} panels (e) and (f), where a beam with OAM $l=0,1$ is simulated under the same conditions as before. Panel (e) shows the phase of the beam at the focus showing a fairly flat spatial profile across the beam and panel (f) shows the intensity profile with a single mode.
\begin{figure}
 \centering
        \includegraphics[width=90mm]{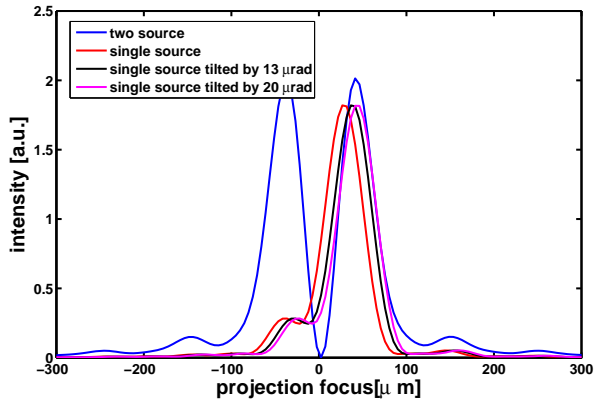}
    \caption{Comparison between a two source ($l=\pm1$, blue) and a single source ($l=1,0$) intensity distribution at the focus of a 75~cm focal length lens for three different values of tilt (no tilt red, 13~$\mu$rad black, and 20~$\mu$rad magenta). The tilt is applied across the entire SLM screen.}
    \label{fig:proj_tilt}
\end{figure}
However, to improve the generation of an identical copy of the individual sources, we are faced with the issue that the superposition of beams with OAM $l=1,0$ does not form the same constructive interference as beams with OAM $l=\pm1$, but forms the constructive interference closer to the optical axis. Imposing a flat wave front tilt on the beam as described in \cite{voelz2011spie}, results in an optimized overlap between single and two source patterns. We found that the tilt applied to the wave front is on the order of 10's of $\mu$rad. Figure \ref{fig:proj_tilt} shows a quantitative comparison between a two source ($l=\pm1$) and a single source ($l=1,0$) intensity distribution at the focus of a 75~cm focal length lens for three different values of tilt. From the figure we can see that the tilt actually guarantees an overlap between one of the two peaks and a single peak distribution. The left shoulder for the $l=1,0$ is almost an order of magnitude lower in intensity than the main peak and thus will have a negligible effect for HHG for example.

\begin{figure}
 \centering
  \includegraphics[width=90mm]{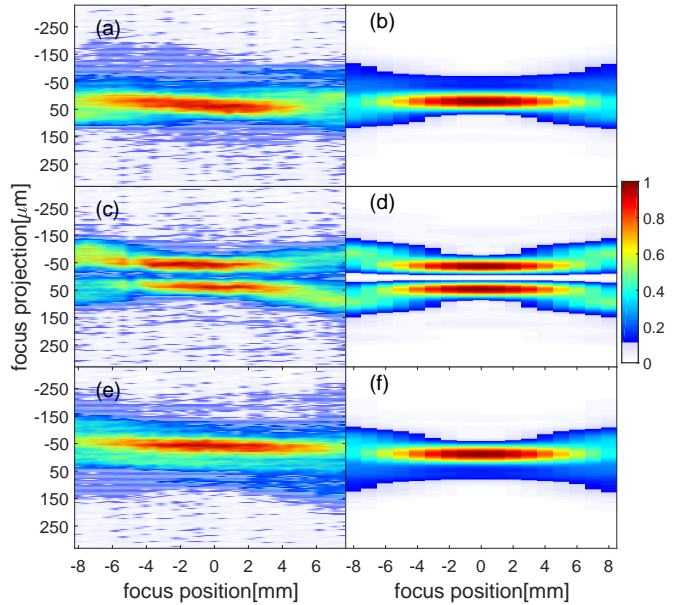}        \caption[Focusing behavior of interfering Laguerre Gaussian beams]{Panel (a) and (b): Projected focus of a laser beam, manipulated by a phase mask that introduces two beams with OAM $l=1,0$. We observe a single focus with a center of mass similar to the center of mass of one of two foci in panel (c) and (d), where we projected the focus of a laser beam, manipulated by a phase mask that introduces two beams with OAM $l=\pm1$. In Panel (e) and (f) we show the projected focus of the other laser focus.}
    \label{fig:focus_LG}
\end{figure}

In Figure \ref{fig:focus_LG}, we compare the mentioned calculations and experimental observations. Imaging of the modes was done with a CCD camera that was moved with a motorized linear stage. In the figure, beam waist calculations of $|U'(x,y)|^2$ are shown as a function of distance around the focus of a 75~cm lens and measure the spot size in the experiment. On the left, we show experimental findings, while on the right we show the theoretical calculations for the same phase manipulations. In panels (c) and (d) we use $l=\pm1$ yielding two foci. Panels (a)-(b) and (e)-(f) where generated using $l=1,0$. The only difference between the two set of panels is an offset applied to the flat phase mask with either $\phi_2=0$ or $\phi_2=\pi$. This also selects the possible condition for destructive interference, besides the sign convention in$l\theta$. On the top two panels the $\phi_2=0$ point is chosen such that the single focus coincides with the top beam in panels (c)-(d) and in the lower two panels the single focus coincide with the bottom beam in (c)-(d).

\section{Molecular alignment}

As seen in the Quantitative Rescattering (QRS) model \cite{le2009PRA}, the harmonic yield of molecules in the molecular frame depends on the angle between the molecular axis and the laser polarization axis. The molecule's axis is given by the molecular frame $z$-axis and the pump and probe laser polarization is defined by the laboratory frame $Z$-axis. In HHG, the dominant ionization channel is the valence shell or the HOMO of the target atom or molecule. The specific symmetries and dynamics of this particular orbital will then be encoded in the yield of higher harmonics. In nitrogen, the HOMO has the symmetry of a $\sigma_g$ orbital, which has an angular density distribution that aligns with the molecular axis and has two nodes perpendicular to the molecular axis. This symmetry would be also visibly recorded on the higher harmonic yield, if the molecule are aligned and rotated in space. This angle dependence is due to the molecule's angular dependent ionization and photorecombination rate, in the molecular frame.
We use an ansatz suggested in \cite{ramakrishna2007PRL} to describe the angle dependent signal of nitrogen as a sum of $\cos^{2n}\theta$-terms and $\sin^2\theta\cos^{2n}\theta$-terms. In our expansion, terms of order $\sin^2\theta\cos^{2n}\theta$ can be expressed as the difference between higher order $\cos^{2n}\theta$-terms, which reduces the expansion to $\cos^{2n}\theta$-terms. By adding an appropriate amount of terms of order $n$ to the expansion, we will be able to describe the correct angle dependence. The angle dependent yield $S(\theta,\omega)$ of a harmonic with energy $\hbar\omega$ in the molecular frame is formulated as
\begin{equation}
\label{eq:mol_frame_n2}
S(\theta,\omega)=\sum_{n}C_n(\omega)\cos^{2n}\theta.
\end{equation}
where the signal in the laboratory frame $S(\theta,\omega)$ is the convolution of the molecular frame signal over the molecular axis distribution. The molecular axis has a distribution function that depends on the non-adiabatic alignment of the molecule.
The molecular axis distribution $\rho$ is defined by
\begin{equation}
\rho(\theta,t)=g_i\frac{e^{E_i/kT}}{Z}|\Psi_i(\theta,t)|^2
\label{eq2}
\end{equation}
where $i=\{J_0,M_0\}$ is the quantum numbers of the involved states, $g$ the nuclear spin state weights, $k$ the Boltzmann constant, $T$ the rotational temperature and $Z$ the partition function. The molecular axis distribution $\rho$ can be interpreted as a probability function of finding the molecule at time $t$ aligned at the angle $\theta$. The distribution can be calculated for the experimentally given laser pulse and gas parameters. \newline
In the laboratory frame, the time dependent signal is defined as an integral over all angles $\theta$
\begin{eqnarray}
S(t)&=&\int\rho(\theta,t)S(\theta)\sin\theta d\theta \\
&=&\sum_{n}C_{n}\int\rho(\theta,t)\cos^{2n}\theta\sin\theta d\theta
\label{eq:hhg_time}
\end{eqnarray}
where the solution to the integral for a particular order $n$ yields
\begin{equation}
\int\rho(\theta,t)\cos^{2n}\theta\sin\theta d\theta=\left<\Psi_i\right|\cos^{2n}\theta\left|\Psi_i\right>(t)
\end{equation}
Each term of the expansion is averaged over the molecular axis distribution.
Using this in the time dependent expansion given through Equation \ref{eq:hhg_time}, the measured signal $S(\omega,t)$ is then 
\begin{equation}
S(\omega,t)=\sum_{n}C_n\left<\Psi_i\right|\cos^{2n}\theta\left|\Psi_i\right>(t).
\label{eq:n2expand}
\end{equation}
The coefficients $C_n$ are complex and follow the equation $C=A+iB$. Changing the parameters in Eq \ref{eq2}, in multiple linear regression fits, will result in the residual being minimized. This results in more confidence on the molecular axis distribution. For the smallest residue, the coefficients $C_n$ can be inserted in the angle-dependent expansion in Equation \ref{eq:mol_frame_n2} and will define the extracted molecular frame harmonic signal $S(\omega,\theta)$. \newline


\section{Experimental Results\label{sec:result}}

\begin{figure}
 \centering
        \includegraphics[width=\columnwidth]{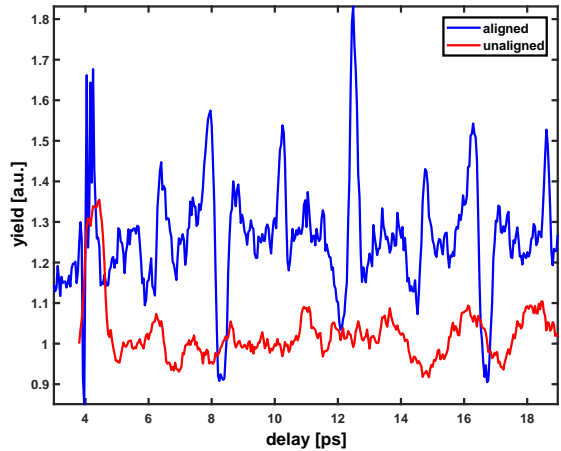}
    \caption[Harmonic yield of single sources]{Harmonic yield as a function of time between pump and probe, generated by a shaped focus to match the focus of the individual ``slits" top and bottom in the experiment.}
    \label{fig:n2_top_bottom}
\end{figure}
\begin{figure}
 \centering
        \includegraphics[width=1.1\columnwidth]{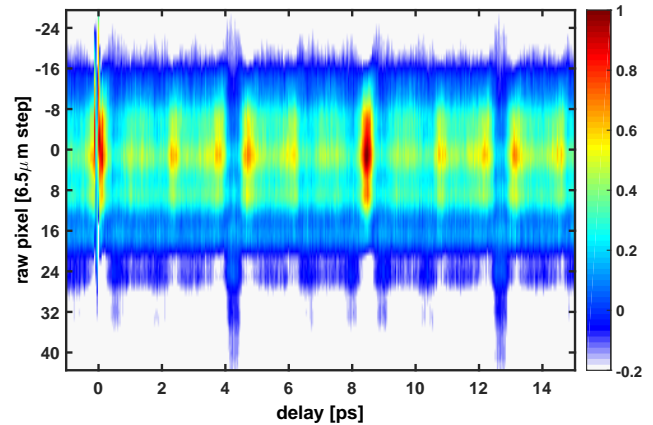}
    \caption[Interference pattern of H19 as a function of time]{Interference pattern of harmonic 19 generated by two intense laser foci. The projection is recorded as a function of time between pump and probe. The fringe spacing is 8 pixels, which equals to $2\pi$ in phase. A shift of 1 pixel equals a phase change of 0.78~rad.}
    \label{fig:n2_fringe19}
\end{figure}
\begin{figure}
 \centering
        \includegraphics[width=0.8\columnwidth]{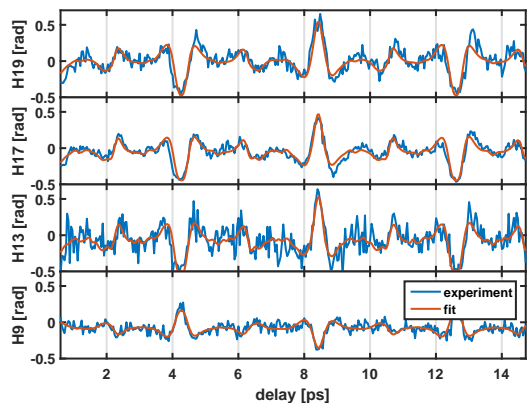}
    \caption[Phase of harmonics in N$_2$ extracted from the double slit]{Phase of harmonic 9, 15 and 19 as a function of time between pump and probe pulses. A fit, based on estimates for the alignment distribution, is shown.}
    \label{fig:n2_phase_fit}
\end{figure}

The experimental yields of harmonics, for alignment with a weak, non-resonant 785~nm pulse and driven by a delayed 785~nm probe pulse, are shown in Figure \ref{fig:n2_top_bottom}. In the blue color, the harmonic yield is given for an experimental conditions, where the pump and the probe pulse were spatially overlapped. For the experimental condition yielding the data in the red color, no second weak field was present. We observe a periodic time dependence for the blue data set. The harmonic yield strongly depends on the time between pump and probe: At 8.3~ps, the yield is reduced to a value of 0.9 and at 12.1~ps, the yield increases to a value of 2. The yields are normalized to their isotropic value, when no alignment beam is present. We observe quarter revivals at 6.1~ps, 10.2~ps and 14.2~ps. 1/8th revivals are visible at 5~ps, 7~ps, 9~ps and 11~ps. The unaligned source, in red, where no spatial overlap is visible between pump and probe in the imaging setup, shows no periodicity in time. We do, however, observe a cross-correlation feature at 4~ps, when pump and probe are incident at the same time. To extract the molecular frame signal, the linear regression is based on the experimental data after the interaction with the pump pulse and the cross-correlation peak does not influence the real physical observations between the unaligned reference source and the aligned second source.
In the shown data, the pump beam was spatially aligned to the top spot of the interferometer focus, but the probe beam was spatially shaped to overlap with either the top or bottom spot of the interferometer, leading to a time dependent yield or a time independent yield.
After the check of the individual sources and that no rotational alignment is visible in the source with no pump pulse, the pattern is changed to the two-source interference mask and the harmonic yield and phase is collected as a function of time between pump and probe. The yield from the aligned source is normalized to the total yield of both sources $I_{total}$.
\begin{equation}
I_{total}=I_1+I_2+2\sqrt{I_1I_2}\cos(2\pi\frac{\delta Y}{\lambda z})
\label{eq:young_intensity}
\end{equation}
where $Y$ is the ordinate of the fringe projection, $I_{1,2}$ the intensities of the harmonic sources, $\lambda$ the wavelength of the light and $z$ the distance to the observation plane.
Integrating over ordinate $Y$, the fringe-angle-integrated yield is equal to $I_{total}=I_1+I_2$. $I_1$ and $I_2$ are identical sources, when no aligning pump beam is present. The isotropic yield of the individual harmonic source is then defined to be $I_{1,2,iso}=I_{total}/2$. The delay dependent yield measurement is normalized to this isotropic value and the intensity of a single source as a function of time can be expressed. To calculate the amplitude of this source, we take the square root of the intensity.
\begin{figure}
 \centering
        \includegraphics[width=0.9\columnwidth]{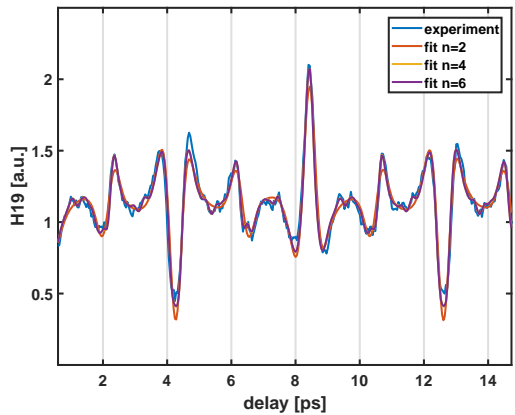}
    \caption[Amplitude measured in the two source experiment in N$_2$]{Amplitude of harmonic 19 recorded as a function of time between pump and probe pulses. The experimental data is fitted to an expansion of varying order.}
    \label{fig:h19_amp_fits}
\end{figure}
In Figure \ref{fig:h19_amp_fits}, the time dependent amplitude of harmonic 19, normalized to the isotropic value, is shown. The experimentally collected data shows an amplitude of harmonic 19 between 0.5 and 2, when normalized to the isotropic value. At a time of 4.1~ps and 12.3~ps, we see a strong anti-alignment dip in the harmonic amplitude, while we observe a maximum in the recorded yield at 8.2~ps. Besides quarter revivals at 2.1~ps and 6.1~ps, smaller revivals in between the quarter revivals are observable, where an oscillation with a strength on the order of 0.1 compared to the isotropic value is present. A fit with the expansion in Equation \ref{eq:n2expand} is performed and higher order terms are added. Smaller features in the delay dependence can only be fitted through the addition of higher order terms with $n=2,3$. Especially, 1/8th revivals at 3 and 5~ps are only fitted with higher order terms. The phase of harmonic 19 as a function of delay is extracted using fast Fourier transformations of the collected Young's double slit fringe pattern. In Figure \ref{fig:n2_fringe19}, the fringe pattern of harmonic 19 is given as a function of delay between pump and probe. The fringes change position most visibly at the times of 4.1~ps and 12.2~ps in the given pattern. With a fringe spacing of 8 pixels, a change in phase of $\pi/4=0.79$~rad for this harmonics is given for each pixel the fringe pattern moves. In the experiment, the observed phase change is on the order of 0.5~rad in the fast Fourier transformation and the equivalent pixel shift is on the order of 2/3 pixel for the shown harmonic, which we can resolve based on the fact that multiple oscillations occur and allow us to have higher resolution sampling of the fringe's movement. Estimates for the resolution can be made in accordance with \cite{bone1986AO}. We perform a series of Fourier transformations for all delays and harmonic orders and extract the phase of individual harmonics as a function of time.
From the measured interference, we obtain a complex valued quantity with $S(\omega,t)=\sqrt{P(\omega,t)}e^{i\phi(\omega,t)}$, in which the imaginary and complex part depend on the phase and amplitude measured in the experiment. To perform a linear regression, the complex number is split into real and imaginary part and two linear regressions are performed., as the equation splits into two linear equations. After the linear regressions, we convert the complex numbers back into amplitude and phase.
In Figure \ref{fig:n2_phase_fit}, we show the time-dependent, measured phases of harmonic 9, 13, 17 and 19. At times of alignment and anti-alignment, the biggest phase offset compared to the reference source is observed. The measured phase of harmonic 9 shows a maximum in phase at a delay of 4.1~ps, while higher order harmonics show a minimum in phase at this delay, as previously reported by \cite{camper2015photonics,soifer2010PRL}. No higher order features, e.g. 1/8th revivals, are present in the time dependent phase of the recorded harmonics, but changes in phase can be measured at the times of quarter and half-revivals. Harmonic 19 shows a variation of up to 1~rad as a function of time to the isotropic value. The shown fits, based on a least square method, show agreement with the measurement, however, do not match the width of the peaks at times of anti-alignment at 4.1~ps and 12.4~ps. The experimental curves show a broader feature in time than the fits can re-produce. 

\section{Angular contributions}

\begin{figure}
 \centering
        \includegraphics[width=\columnwidth]{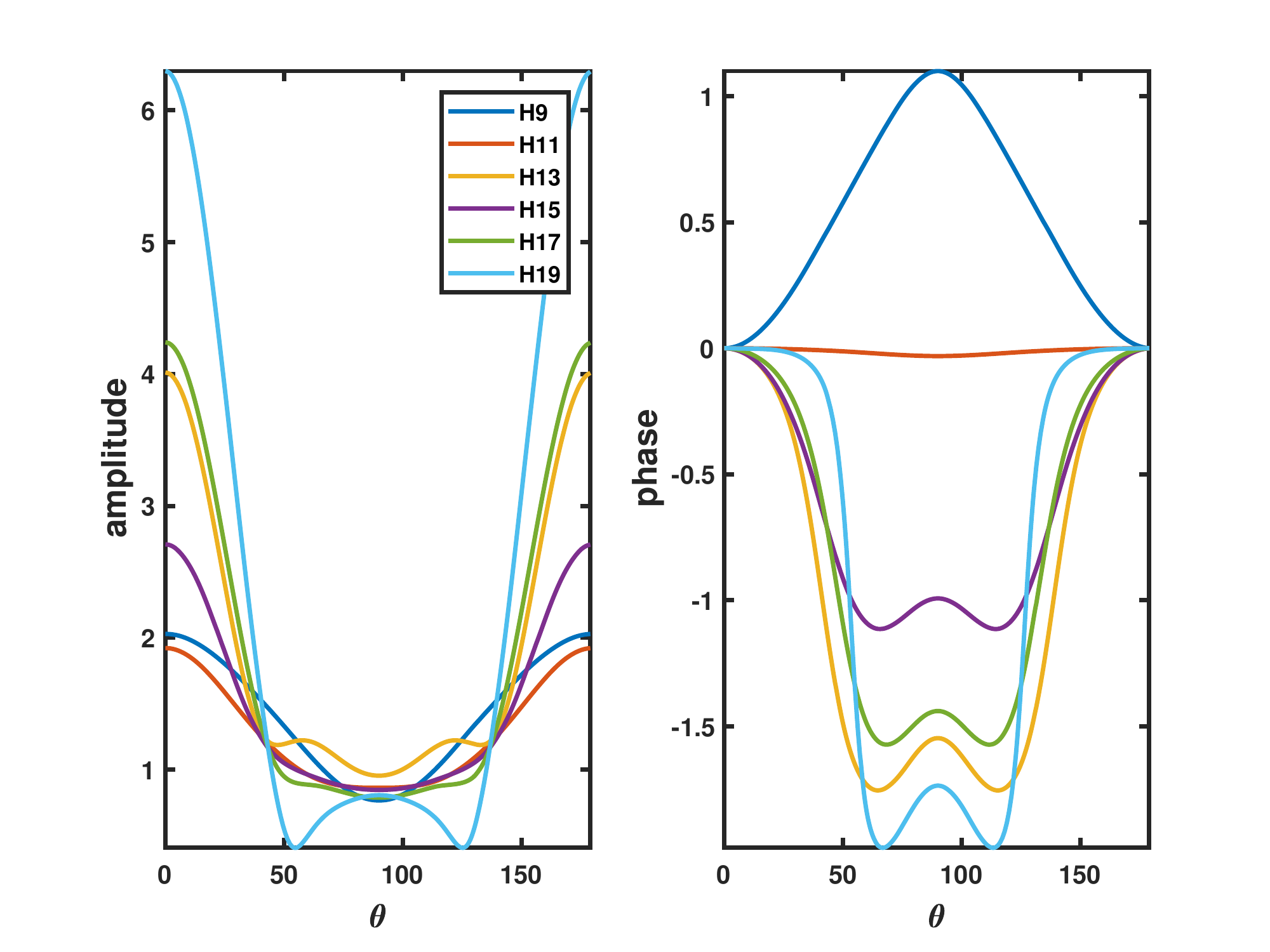}
    \caption[Molecular frame phase and amplitude in N$_2$]{Phase and amplitude of the harmonic emission in the molecular frame as a function of angle between probe polarization and molecular axis for measured harmonics.}
    \label{fig:n2_mf_phase}
\end{figure}
\begin{figure}
\centering
\includegraphics[width=\columnwidth]{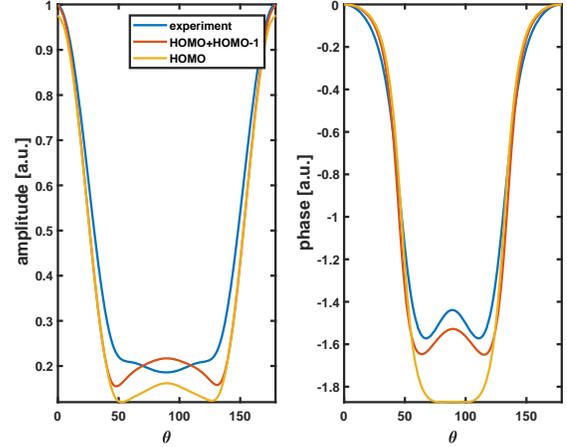}
\caption[Comparison molecular frame dipole to calculated signal]{Harmonic amplitude and phase of H17 as a function of angle $\theta$. The experimentally retrieved angle dependent phase and amplitude is compared to harmonic emission calculated by the factorization of the harmonic yield in ionization rate, given by MO-ADK \cite{le2009jopb}, the phase difference due to the difference in ionization potential and the complex photoionization cross section}
\label{fig:n2_pics_17}
\end{figure}

The angle dependent amplitude and phase in the molecular Eq. \ref{eq:mol_frame_n2}frame is plotted in Figures \ref{fig:n2_mf_phase}. To calculate the quantity $S(\theta,\omega)$ we used the extracted coefficients $C_n$, as detailed in Equation \ref{eq:n2expand}. As mentioned before, both $S(\theta,\omega)$ and $C_n$ are complex valued. We observe in the left panel an angle dependent amplitude that has a maximum at 0$^\circ$ and a local minimum at 90$^\circ$. As the harmonic order increases, so does the maximum harmonic amplitude. For the 9th and 11th order, we observe a maximum of 2, while for H17, we observe a maximum of 4 and for H19 a maximum of 6, normalized to its original isotropic value. The extracted phase of the recorded harmonics shows a similar behavior. As the harmonic order increases, so does the absolute change in phase between 0 and 90$^\circ$ for the different orders. In the angle-dependent phase, we can observe another feature. At 90$^\circ$, a local maximum in the extracted phase values is visible. Harmonic 9 shows  the opposite angle-dependent behavior, as mentioned earlier in the delay dependent phase measurements. Harmonic order 19 is showing a strong angle dependence in amplitude with a angle-resolved amplitude of six times the isotropic value at an angle of $\theta=0^\circ$. A phase difference of 1.6~rad between harmonic emission at $\theta=0^\circ$ and $\theta=90^\circ$ is visible. This feature can be explained with the shape resonance in the photoionization cross section of the HOMO at the particular photon energy. The residuals for the imaginary and real part of the time dependent signal is reducing with order, when higher order terms are being added. The addition of terms of order $2n=6$ do not improve the fit to the experimental data. Orders $2n=4$ and $2n=6$ can predict smaller fractional revivals and do not differ drastically for the given temperature and pulse intensities, so that a fit to $2n=6$ does not improve the delay dependent fit to the experimental data.
In Figure \ref{fig:n2_pics_17}, the extracted phase and amplitude of harmonic 17 is given as a function of angle between the molecular axis and the driving laser polarization. We compare the angle-dependence to the angle dependence calculated by the factorization in QRS:
\begin{eqnarray}
D_{total}(\omega,\theta)=&&N_{HOMO}(\theta))^{1/2}d_{HOMO}(\omega,\theta)\\
&+&(N_{HOMO-1}(\theta))^{1/2}d_{HOMO-1}(\omega,\theta)e^{i\Delta\eta}
\end{eqnarray}
where $D_{total}(\omega,\theta)$ is the coherent sum of harmonic dipoles from HOMO and HOMO-1 with the ionization potential difference of HOMO and HOMO-1 of 1.3~eV. The ionization rates $N(\theta)$ are given by a theoretical ionization calculation of HOMO and HOMO-1 by MO-ADK theory extracted from \cite{le2009jopb} and $d(\omega,\theta)$ is supplied from \cite{jin2012PRA} with $d=\sqrt{\sigma}e^{i\phi}$. For the ionization rates of HOMO and HOMO-1, we use a ratio of 5:1 for the preferential ionization of HOMO over HOMO-1 at 90$^\circ$. The angle dependent ionization rate for HOMO has a ratio of 9:1 for ionizing parallel to the molecular axis compared to ionizing perpendicular to the molecular axis. A phase difference given by the classical action of the electron in the continuum is given by the ionization potential difference between the two molecular orbitals and is accounted for by $e^{i\Delta\eta}$. The theoretical PICS calculation then allows us to calculate the harmonic dipole as a product of the given complex-valued amplitudes of ionization rate, photoionization cross section and electron wave packet. 
The harmonic dipoles are plotted for harmonic order 17 in Fig. \ref{fig:n2_pics_17}. The persistent feature, visible in the angle dependent phase measurement at 90$^\circ$ can be explained only by using non-vanishing probabilities of HOMO-1 to the total harmonic dipole. We can match the retrieved angle dependent phase of harmonic 15 and 17. Harmonic 19 can be explained by HOMO only, but shows better agreement with the experiment, when a portion of HOMO-1 is added to the calculated total dipole.
Here we assumed an ionization rate similar to the MO-ADK model rate given by publication \cite{le2009jopb}, where the ratio of parallel to perpendicular ionization rate of HOMO is given with 10:1. However, rates of 3.3:1 \cite{pavivcic2007PRL} and 4.5:1 \cite{litvinyuk2003PRL} have been measured. Since we did not measure the angle-dependent single-ionization yield in our experiment, we used the MO-ADK rate.
\section{Concluding remarks}
Using OAM to generate two tightly spaced foci has proven a reliable and stable interferometric setup. Since the two beams with opposite OAM are sampled over the whole SLM surface, the interferometer is very robust to external perturbations. Furthermore, thanks to extra degrees of control introduced by the SLM aligning the experiment is very trivial compared with other multi-beam setups. Using a combination of OAM and a flat phase $l=1,0$ allows us to experimentally verify the proper overlap between the two sources and the aligning pump pulses. Therefore, these new sets of multi-beam experiments, controlled by a SLM, provide a very viable platform for homodyne measurements where the local oscillator is the XUV beam where no pump is present.

Our experimental findings fit well with previous experimental findings. As we did before in a phase matching dependence study\cite{Tros2017PRA}, we observe a strong signature from the lower lying orbital HOMO-1. This time, this features is very significant in the retrieved angular distributions. This new sensitivity is due in part to the fact that we are using a homodyne measurement. To the best of our knowledge this is the first time that a measurement has been performed to retrieve the photoionization phase of the N$_2$ HOMO-1 orbital.\\
In this experiment, we report HHG from HOMO and HOMO-1 for low order harmonics, which match the characteristic features of the photoionization cross section in phase and amplitude. In previous work \cite{mcfarland2009PRA}, features from HOMO-1 in nitrogen were restricted to cut-off harmonics with photon energies of harmonic order 35 and higher. Here we report low order harmonics in nitrogen that are generated from HOMO-1.
\section{Acknowledgements}
This work was supported by the Chemical Sciences, Geosciences, and Biosciences Division, Office of Basic Energy Sciences, Office of Science, U.S. Department of Energy (DOE) under Grant No. DE-FG02-86ER13491. C.A. T-H was partially funded by the Chemical Sciences, Geosciences, and Biosciences Division, Office of Basic Energy Sciences, Office of Science, U.S. Department of Energy (DOE) Grant No.DE-SC0019098


\nocite{*}
\bibliography{Biblio}

\end{document}